\documentclass[%
reprint,
superscriptaddress,
amsmath,amssymb,
aps,
]{revtex4-2}
\usepackage{graphicx}
\usepackage{dcolumn}
\usepackage{bm}
\usepackage{xcolor}
\usepackage{multirow}
\usepackage{amsmath,amssymb,amsfonts}
\usepackage{array,tabularx}
\usepackage{booktabs}                 
\usepackage[colorlinks,linkcolor=blue, urlcolor=blue, anchorcolor=blue, citecolor=blue]{hyperref}
\usepackage{hyperref}
\begin{document}
	\maxdeadcycles=1000
	\preprint{APS}

\title{Quantum phase transitions of the Cavity Heisenberg spin-chain}



\author{Lv-Ting Gong}
\affiliation{College of Physics and Electronic Engineering, Northwest Normal University, Lanzhou, 730070, China}
\author{Shao-Fan He}
\affiliation{College of Physics and Electronic Engineering, Northwest Normal University, Lanzhou, 730070, China}
\author{Fu-Quan Dou}
\email{doufq@nwnu.edu.cn}
\affiliation{College of Physics and Electronic Engineering, Northwest Normal University, Lanzhou, 730070, China}
\affiliation{Gansu Provincial Research Center for Basic Disciplines of Quantum Physics, Lanzhou, 730000, China}

\begin{abstract}
The interplay between quantum criticality and ergodicity breaking constitutes a central challenge in complex quantum many-body systems. Here, we investigate ground-state quantum phase transitions (QPTs) and excited-state quantum phase transitions (ESQPTs), as well as ergodic-nonergodic transition in the cavity Heisenberg spin-chain (CHS) model. By combining quantum information measures with semiclassical fixed-point analysis, we identify a deformed phase that interpolates between the normal and superradiant phases, featuring a logarithmic nonanalyticity in the density of states (DoS) and two additional jump discontinuities. We further elucidate the spectral-structure mechanism underlying the connection between ESQPTs and the ergodic--nonergodic transition (ENET) via level statistics, participation ratios, and multifractal analysis. Our results provide a general framework for characterizing phase structures and spectral features in light--matter quantum many-body systems.
\end{abstract}
\maketitle
\section{INTRODUCTION} \label{section1}
Critical phenomena in quantum many-body systems represent a fundamental theme in modern quantum physics~\cite{Sachdev_2011}. Among them, quantum phase transitions (QPTs) are prototypical examples that occur at zero temperature, where quantum fluctuations prevail over thermal ones~\cite{Baumann2010}. QPTs describe abrupt and nonanalytic changes in the ground-state properties as the control parameter of a system is varied, and the critical value at which the transition occurs is referred to as the quantum critical point~\cite{Sachdev_2011,PhysRevA.94.023835,MatthiasVojta2003}. These transitions are intrinsically associated with global reorganizations of the Hilbert space structure, often involving spontaneous symmetry breaking or changes in topological order~\cite{RevModPhys.69.315,RevModPhys.89.041004,KITAEV20062}. QPTs provide deep insights into universal scaling laws, renormalization group flows, and the emergence of low-energy excitations~\cite{Sachdev_2011} and offer a unified theoretical framework for classification of quantum phases of matter characterized by either symmetry breaking or topological order~\cite{Sachdev_2011,KITAEV20062,RevModPhys.89.041004}. The rapid development of quantum information science~\cite{Nielsen2010,Sachdev_2011,PhysRevB.101.115142} has further stimulated interest in the interplay between QPTs and quantum correlations, including entanglement~\cite{PhysRevLett.92.073602,PhysRevLett.109.017202,PhysRevLett.115.046603,RevModPhys.80.517,Osterloh2002}, mutual information~\cite{PhysRevLett.92.067902,PhysRevA.83.012327,PhysRevLett.105.030501}, and quantum chaos~\cite{PhysRevLett.90.044101,PhysRevE.83.046208}.

In recent decades, significant attention has been directed towards critical phenomena occurring in excited states, which are known as excited-state quantum phase transitions (ESQPTs)~\cite{PhysRevE.66.016217,PhysRevLett.95.050402,Stransky2014,Cejnar2021,CAPRIO20081106}. ESQPTs are characterized by non-analytic behaviors in individual excited states as functions of a controlling parameter. The occurrence of an ESQPT is closely related to the nonanalytic behavior of the energy-level flow and the density of states (DoS) of the system within the mean-field approximation or the large-$N$ limit~\cite{Cejnar2021,STRANSKY20162637}. Specifically, the energy at which the DoS diverges is identified as the critical energy of the ESQPT. As a natural extension of ground-state quantum phase transitions, ESQPTs have been found to emerge in a wide variety of quantum systems, including the Lipkin-Meshkov-Glick model~\cite{PhysRevLett.99.050402,PhysRevB.106.024311,PhysRevE.104.034119,PhysRevLett.95.050402}, the Dicke model~\cite{PhysRevA.107.043706,PhysRevE.94.022209,PhysRevE.83.046208,PhysRevA.89.032101,PhysRevA.98.043805}, the Jaynes-Cummings model~\cite{PhysRevA.108.033710}, Rabi models~\cite{PhysRevA.108.033710,PhysRevA.94.023835}, the Kerr nonlinear
oscillator~\cite{ChCarlos2023,PhysRevE.96.012121}, the interacting boson model~\cite{PhysRevC.99.064323} and spinor Bose-Einstein condensates (BECs) ~\cite{PhysRevE.110.064112,PhysRevA.107.033307,PhysRevLett.126.230602,PhysRevResearch.3.043215,PhysRevResearch.5.013087,PhysRevA.107.033307}. It is known that ESQPTs can significantly affect quantum dynamics after a quench~\cite{PhysRevA.83.033802,PhysRevA.104.053722,PhysRevA.94.012113,PhysRevE.103.032109,PhysRevE.92.012101,PhysRevA.100.022118,PhysRevA.78.060102,PhysRevA.100.062113,PhysRevA.103.032213,PhysRevA.80.032111}, lead to localized eigenstates~\cite{PhysRevA.92.050101,PhysRevA.94.012113}, accelerate the time evolution of the
system~\cite{PhysRevA.98.013836,PhysRevLett.123.160401,PhysRevE.101.010202,PhysRevE.110.064112}, and facilitate the design of quantum cat states~\cite{PhysRevA.105.052204}. Furthermore, ESQPTs have been closely linked to the onset of quantum chaos~\cite{PhysRevE.104.L062202,PhysRevE.83.046208,Corps2022,PhysRevE.94.012140}, various types of dynamical quantum phase transitions~\cite{PhysRevB.106.024311,PhysRevB.107.094307,PhysRevLett.130.100402}, and thermal phase transitions~\cite{PhysRevE.96.012121}. Experimentally, some of these quantum phase transitions have been observed in BECs~\cite{PhysRevLett.126.230602} and quantum cavity systems~\cite{PhysRevResearch.3.043215}. Moreover, the rapid advancement of cavity quantum electrodynamics technology has further accelerated research on systems involving light-matter interactions~\cite{FriskKockum2019,Landig2016,Zhou2019}.

The cavity Heisenberg spin-chain (CHS) model describes the coupling between a Heisenberg spin chain and a single-mode radiation field in a cavity, and it can equivalently be viewed as a class of light-matter interacting systems~\cite{PhysRevA.106.032212,Sun2025}. The Dicke model--a special case of CHS--plays a key role in the study of light-matter interactions, especially quantum phase transitions. 
 In the thermodynamic limit (\(N\to\infty\)), the system undergoes a second-order quantum phase transition: as the light--matter coupling reaches a critical value, the ground state evolves from the normal phase (NP) to the superradiant phase (SP)~\cite{hepp1973superradiant,PhysRevA.7.831}. Experimental demonstrations of tunable light-matter coupling have since provided versatile platforms for probing quantum critical phenomena, with realizations across diverse quantum systems including superconducting qubits~\cite{Mezzacapo2014,PhysRevA.69.062320,PhysRevLett.105.263603}, BECs in optical lattices~\cite{Klinder2015,doi:10.1126/science.1083171,Baumann2010,PhysRevLett.107.140402,PhysRevLett.105.043001}, cavity-assisted Raman transitions~\cite{PhysRevLett.113.020408,PhysRevA.75.013804}, and unitary Fermi gases~\cite{doi:10.1126/science.abd4385,Helson2023}. Notably, the experimental observation of the superradiant phase transition in BECs has further spurred systematic investigations into Dicke-like Hamiltonians and their extended formulations~\cite{Baumann2010,PhysRevLett.104.130401,PhysRevLett.107.140402,Pilar2020thermodynamicsof}. Beyond their experimental relevance, these models also exhibit a relatively simple algebraic structure and are closely intertwined with fundamental quantum phenomena including entanglement~\cite{PhysRevA.106.032212,Sun2025,PhysRevLett.92.073602} and quantum chaos~\cite{PhysRevA.98.043805,PhysRevE.67.066203,PhysRevLett.109.179301}. This combination of structural simplicity and rich quantum behavior makes these models particularly valuable even for finite-\(N\) systems~\cite{Vidal2006}.
 As an extension of the Dicke model, the CHS model has also been explored for quantum battery applications, with such explorations revealing that spin-spin interactions and cavity-spin coupling both exert a substantial influence on charging performance and bear close links to the critical behavior of QPTs~\cite{PhysRevA.106.032212,Sun2025}.
 Recent studies have shown that $XXZ$-type spin--spin interactions can substantially reshape the ground-state phase structure, giving rise to a quasi--long-range-ordered $XY$-type phase and resulting in more intricate many-body phase diagrams~\cite{10.1103/z8gv7yyk,PhysRevA.104.013303}.
 However, within more general anisotropic interaction frameworks, a unified, systematic characterization of the full phase structure and precise phase boundaries is yet to be fully established---most notably for the combination of analytical theory and numerical simulations. This limitation impedes in-depth insight into the corresponding energy operating regimes and their underlying physical mechanisms. 

In this work, we report a systematic characterization of ground-state and excited-state criticality along with ergodic-nonergodic behavior in the CHS model. We first characterize ground-state QPTs via quantum-information-inspired diagnostics. Guided by the order parameter \(\langle J_z\rangle\), scaled mean photon number, and inverse participation ratio (IPR), we identify an intermediate deformed phase that separates the normal and superradiant phases, whose existence is further corroborated by semiclassical fixed-point analysis. We then investigate the excitation spectrum and its statistical properties--including the adjacent-level-spacing ratio, DoS, participation ratios, and multifractal dimensions to uncover spectral fingerprints of ESQPTs and the ergodic-nonergodic transition (ENET). Building on these findings, we analyze quench dynamics by expanding the initial state in the eigenbasis of the post-quench Hamiltonian and tracking its energy-space distribution, demonstrating that ESQPT induced spectral restructuring near the critical energy directly governs the ensuing nonequilibrium response.

This rest of paper is organized as follows. In Sec.~\ref{section2}, we introduce the model Hamiltonian. In Sec.~\ref{section3}, we discuss the QPTs and characterize them using quantum information measures alongside semiclassical analytical methods. Sec.~\ref{section4} is dedicated to the characterization of ESQPTs via the DoS and spectral statistics. In Sec.~\ref{section5}, we investigate the ENET by employing participation ratios and multifractal dimensions. Finally, we give a summary in~Sec.~\ref{section6}.
\section{MODEL} \label{section2}
We consider the CHS model, described by the Hamiltonian~\cite{PhysRevA.106.032212,Sun2025} (we set \(\hbar=1\) throughout):
\begin{align}\label{HamC1}
\hat{H} = &\, \hat{H}_{XYZ} + \omega_c \hat{a}^\dagger \hat{a}+\frac{g_1}{\sqrt{N}} \sum_{i=1}^N \hat{\sigma}_i^x (\hat{a}^\dagger + \hat{a})
+ \frac{\omega_a}{2} \sum_{i=1}^N \hat{\sigma}_i^z,
\end{align}
with
\begin{align}\label{HamC2}
\hat{H}_{XYZ} = &\, \frac{\omega_a g_2}{N} \sum_{i<j}^{N}[(1+\gamma) \hat{\sigma}_{i}^{x} \hat{\sigma}_{j}^{x}
+ (1-\gamma) \hat{\sigma}_{i}^{y} \hat{\sigma}_{j}^{y}+ \Delta \hat{\sigma}_{i}^{z} \hat{\sigma}_{j}^{z}].
\end{align}
Here, $\hat{\sigma}_i^\alpha$ is the Pauli operators of the $i$th site, and \( \hat{J}_\alpha = \frac{\hbar}{2} \sum_{i=1}^{N} \hat{\sigma}_i^\alpha \, (\alpha \in \{x, y, z\}) \). \( \hat{a} \) (\( \hat{a}^\dagger \)) annihilates (creates) a photon in the cavity with frequency \(\omega_c\), while the spin-cavity coupling strength is characterized by the dimensionless parameter \( g_1 \), and \( \omega_a \) denotes the spin frequency. The coefficients \( \gamma \) and \( \Delta \) characterize the anisotropy of the system. The conserved pseudospin operator \(\hat{\mathbf{J}}^2 = \hat{J_x}^2 + \hat{J_y}^2 + \hat{J_z}^2 \) commutes with the Hamiltonian, $\left[ \hat{H}, \hat{\mathbf{J}}^2 \right]=0$, which allows the Hilbert space to be decomposed into symmetry sectors labeled by the eigenvalues $j(j+1)$. In the subspace with total spin \( j = N/2 \), the Hamiltonian \( \hat{H} \) reduces to
\begin{align}\label{HamC3}
\hat{H} = &\, \omega_c \hat{a}^\dagger \hat{a} + \frac{\omega_a g_2}{N} \Big[ \hat{J}_+ \hat{J}_- + \hat{J}_- \hat{J}_+
+ \gamma (\hat{J}_+^2 + \hat{J}_-^2) + 2\Delta \hat{J}_z^2 \notag \\
&- \frac{N}{2}(2 + \Delta) \Big]
+ \frac{g_1}{\sqrt{N}} (\hat{J}_+ + \hat{J}_-) (\hat{a}^\dagger + \hat{a})
+ \omega_a \hat{J}_z.
\end{align}

To gain physical insight into the CHS model, it is instructive to examine its classical (semiclassical) limit, as has been extensively demonstrated for the Dicke model~\cite{PhysRevA.89.032101,PhysRevA.83.033802,PhysRevA.98.043805,PhysRevA.91.033819,PhysRevA.98.043805,HerreraRomero2022}. The Hamiltonian~(\ref{HamC3}) involves a finite number of degrees of freedom, and its thermodynamic limit ($j \rightarrow \infty$) corresponds to the semiclassical limit ($\hbar \rightarrow 0$). To derive the classical Hamiltonian, the bosonic annihilation and creation operators ($\hat{a}$, $\hat{a}^\dagger$) are replaced by the canonical variables of a classical harmonic oscillator under the condition $m\omega = 1$, i.e.,
\(2\hat{a} \rightarrow \hat{q} + i\hat{p}, \quad 2\hat{a}^\dagger \rightarrow \hat{q} - i\hat{p},\) where $\hat{q}$ and $\hat{p}$ represent the quantum mechanical position and momentum operators, respectively. Meanwhile, the pseudospin operators $\hat{J}_i$ are mapped onto classical angular momentum components ($J_i \rightarrow j_\alpha)$, where $j_\alpha (\alpha=x,y,z)$ denotes the total spin magnitude, $j_z$ is its projection along the $z$-axis, and $j_x = \sqrt{j^2 - j_z^2}\cos\varphi$, with the azimuthal angle $\varphi$. Assuming conservation of total spin magnitude, one spin component can be expressed in terms of the others via the identity $\hat{J}_z^{2} = j(j+1)\hat{I} - \hat{J}_x^{2} - \hat{J}_y^{2}$, where $\hat{I}$ is the identity operator. When the quantum operators $\hat{q}$ and $\hat{p}$ are considered as classical continuous variables ($q$, $p$), the semiclassical Hamiltonian follows:
\begin{align}\label{HamCl}
H_{cl} &= \frac{\omega_c}{2}(q^2 + p^2) + \omega_a j_z + 2 g_1 q \left(\frac{j^2 - j_z^2}{j}\right)^{1/2} \cos \varphi \nonumber \\
&\quad + \frac{\omega_a g_2 (j^2 - j_z^2)}{j} \left[\gamma (\cos^2 \varphi - \sin^2 \varphi) + 1\right] \nonumber\\
&+ \frac{\omega_a g_2 \Delta j_z^2}{j}.
\end{align}
\section{QUANTUM PHASE TRANSITIONS} \label{section3}
We effectively characterize QPTs via a combination of numerical simulations and analytical methods. Tracking key observables-magnetization, photon number, and localization measures, we detect pronounced changes across phase boundaries that distinctly signal ground-state structural transformations. Complementing these observables, we carry out a semiclassical analysis based on the fixed-point structure of the classical Hamiltonian, affording geometric insights into the phase structure and critical behavior.
\subsection{Signatures of ground-state QPTs in physical observables}
\begin{figure*}[htbp]
\centering
\includegraphics[width=0.85\textwidth]{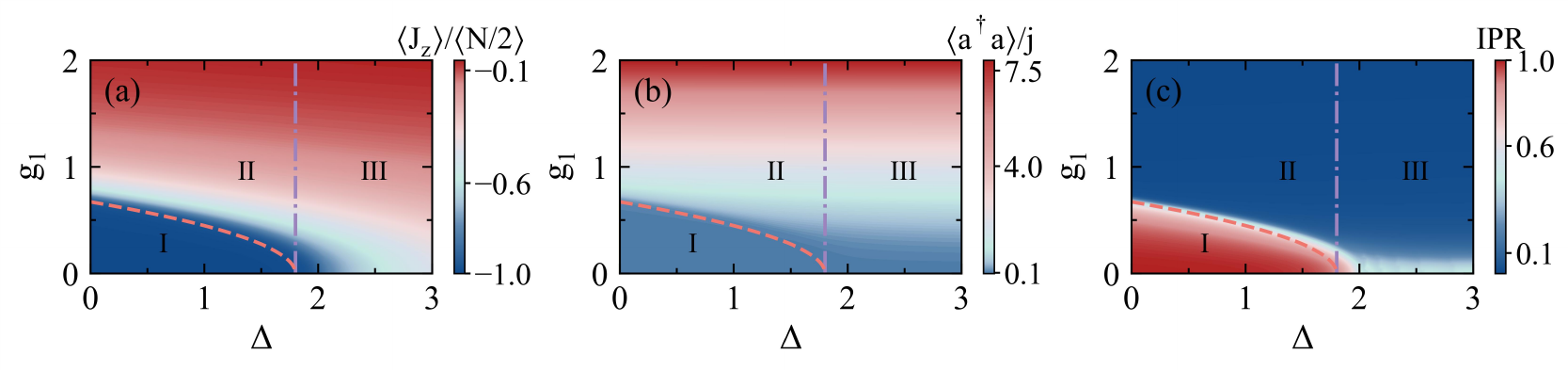}
\caption{\label{fig.1}(a) The order parameter \(\langle J_z \rangle / (N/2)\) distinguishes the NP with \(|\langle J_z \rangle / (N/2)| = 1\) from the SP with \(0 < |\langle J_z \rangle / (N/2)| < 1\).
(b) The scaled mean photon number \(\langle \hat{a}^\dagger \hat{a} \rangle / j\) is nearly zero in the NP and becomes finite in the SP, showing a continuous change across the phase boundary.
(c) The ground-state IPR approaches $1$ in the NP (localized) and tends toward $0$ in the SP (delocalized). 
The red and yellow dashed lines represent the critical boundaries obtained from the auxiliary function given in Eq.(~\ref{f}), which distinguish the NP, SP, and the intermediate phase. Parameters: \(N = 20\), \(g_2 = 0.5\), \(\gamma = -0.2\), and \(n_{\text{max}} = 100\).}
\end{figure*}

QPTs in the ground state can be identified through distinct signatures in measurable physical observables. In this section, we analyze several representative quantities to characterize and locate the transition points. Figure~\ref{fig.1}(a) presents the phase diagram obtained from the order parameter \( \langle J_z \rangle / (N/2) \). The system resides in the NP when \( \left| \langle J_z \rangle / (N/2) \right| = 1 \), and enters the SP for \( 0 < \left| \langle J_z \rangle / (N/2) \right| < 1 \). 
The scaled mean photon number \(\langle \hat{a}^\dagger \hat{a} \rangle/j\) is shown in Fig.~\ref{fig.1}(b). It remains nearly zero in the NP and becomes finite in the SP, increasing continuously across the transition~\cite{PhysRevE.67.066203,PhysRevA.78.023634}. For clarity, we partition the parameter space into regions~I--III. The corresponding critical lines and their theoretical origin will be established through the semiclassical analysis presented below.
In regions~I and II, the system evolves from the NP to the SP as the coupling parameter $g_1$ increases. Interestingly, in region~III, the results shown in Fig.~\ref{fig.1}(a) display features consistent with the SP, whereas the observable presented in Fig.~\ref{fig.1}(b) retains characteristics of the NP.

To further probe the nature of the ground state, we employ the inverse participation ratio (IPR), a standard metric for wavefunction localization~\cite{Edwards1972,PhysRevResearch.2.043395}. For an eigenstate \(|\psi\rangle = \sum_j \psi_j |j\rangle\) in a Hilbert space of dimension \(N_D\), the IPR is defined as
\begin{equation}\label{IPR}
\text{IPR} = \sum_{j=1}^{N_D} |\psi_j|^4.
\end{equation}
As illustrated in Fig.~\ref{fig.1}(c), the IPR approaches unity in the NP, indicating strong localization of the ground state in the chosen basis. In contrast, in region~III, the IPR follows the trend observed in the SP, indicating a delocalized ground-state wave function.

A natural question arises: could region~III, where conflicting signatures are observed, correspond to an unconventional phase that eludes the standard classification into NP or SP? To address this, we next employ a semiclassical fixed-point analysis to investigate the underlying phase structure and clarify the nature of this anomalous regime.
\subsection{Semiclassical analysis}
Fixed-point equations derived from the classical Hamiltonian play a crucial role in elucidating the phase structure and critical phenomena of the system. They represent semiclassical steady-state conditions, in which the classical variables are time-independent. The number, location, and stability of the fixed points directly characterize the quantum phases of the system and determine the locations of the associated phase transitions, including both QPTs and ESQPTs. Notably, bifurcations or qualitative changes in the fixed-point landscape often signal the emergence of critical behavior. This semiclassical approach offers not only an intuitive perspective on the underlying dynamics but also a powerful framework for guiding numerical simulations and interpreting physical observables. To analyze the fixed points and their stability, we employ the classical equations of motion derived from Hamilton's equations, which govern the time evolution of the canonical variables. The semiclassical equations of motion are as follows,
\begin{align}\label{HamC6}
\frac{d q}{dt}& = \omega_c p,
\end{align}
\begin{align}\label{HamC7}
\frac{dp}{dt}& = -\omega_c q - 2 g_1 \left(\frac{j^2 - j_z^2}{j}\right)^{\frac{1}{2}} \cos \phi,
\end{align}
\begin{align}\label{HamC8}
\frac{d\phi}{dt}& = \omega_a - 2 g_1 q \frac{j_z}{j} \left(\frac{j^2 - j_z^2}{j}\right)^{-\frac{1}{2}} \cos \phi \nonumber \\
& - \frac{2 \omega_a g_2 j_z}{j} \left(\gamma (\cos^2 \phi - \sin^2 \phi) +1\right)
 + \frac{2 \omega_a g_2 \Delta j_z}{j} ,
\end{align}
\begin{align}\label{HamC9}
\frac{dj_z}{dt}& = 2 g_1 q \left(\frac{j^2 - j_z^2}{j}\right)^{\frac{1}{2}} \sin \phi + 2 \omega_a g_2 \gamma \frac{j^2 - j_z^2}{j} \sin(2\phi).
\end{align}
The fixed points of the Hamiltonian (\ref{HamCl}) can be found when all four derivatives simultaneously equal zero. For any given cavity-spin coupling strength \( g_1 \) and spin-spin interaction, two fixed points are obtained:
\begin{align}\label{Point1}
\{q,~p,~j_z\} = \{0,~0,~\pm j\},
\end{align}
The states with~\( j_z = \pm j\)~correspond to the north and south poles of the pseudospin sphere and are independent of the azimuthal angle. Evaluating the Hamiltonian (\ref{HamCl}) at the preceding fixed points yields energy values of~\( \mathcal{E}_\pm = g_2 \Delta \pm 1\).~The nature depends on the auxiliary parameter~\(f_{z \alpha},~\alpha \in \{x,~y\}\)~and the difference between interactions among different spins,~i.e.,~\(\delta \eta_{z \alpha} = \eta_z - \eta_\alpha\) (here $\eta_z={2 \omega_a g_2\Delta}$, $\eta_x={2 \omega_a g_2(1+\gamma)}$, and $\eta_y={2 \omega_a g_2(1-\gamma)}$),
\begin{align}\label{f}
f_{z x} = \frac{4 g_1^2 + \omega_c \delta \eta_{z x}}{\omega_a \omega_c},~f_{z y} = \frac{\delta \eta_{z y}}{\omega_a}.
\end{align}
In particular, the condition \(f_{zx}=1\) marks the onset of the instability of the south-pole fixed point and therefore defines the critical spin--cavity coupling
\begin{equation}\label{g1c}
g_{1c}=\frac{\sqrt{\omega_a\omega_c}}{2}\sqrt{1-2g_2(\Delta-1-\gamma)}.
\end{equation}
Accordingly, the fixed point \(\{q,p,j_z\}=\{0,0,j\}\) is unstable for any value of the auxiliary parameters, whereas the fixed point \(\{q,p,j_z\}=\{0,0,-j\}\) is stable before the threshold is reached and becomes unstable once \(f_{zx}\ge1\). When \(f_{zx}\ge1\) and \(\delta\eta_{zx}<\omega_a\), two degenerate stable fixed points emerge,
\begin{align}\label{Point2}
\{q,~p,~j_z,~\phi\} = \{q_s,~0,~-j f_{z x}^{-1},\pi\}, \{-q_s,~0,~-j f_{z x}^{-1},~0\}
\end{align}
with~\(q_s = 2 g_1 \sqrt{j - j f_{z x}^{-2}} / \omega_c\).~By analyzing the auxiliary function in Eq.~(\ref{f}), we find that when~\(f_{z y} \geq 1\), the nature of the fixed points are independent of~\( g_1 \) (\(f_{z y} = \frac{\delta \eta_{z y}}{\omega_a }\)). Thus, we obtain two new fixed points
\begin{align}\label{Point3}
\{q,~p,~j_z,~\phi\} = \{0,~0,~-j f_{z y}^{-1},~\pm \frac{\pi}{2}\}.
\end{align}
\begin{figure}[htbp]
\centering
\includegraphics[width=0.38\textwidth]{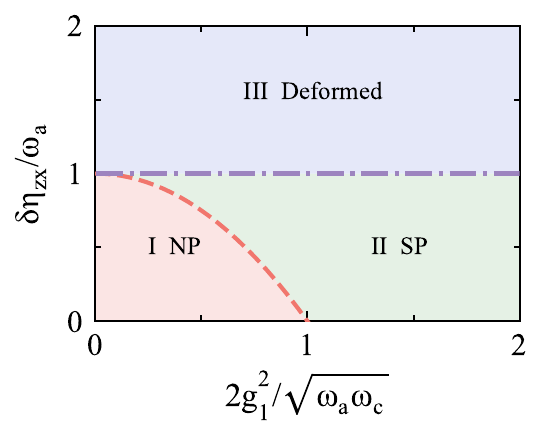}
\caption{\label{fig.2} Fixed point phase diagram of the classical Hamiltonian~(\ref{HamCl}).
The parameter space is divided into three regions according to the number and stability of the stationary points. Region~I corresponds to the NP. Region~II, defined by $f_{zx}\geq1$ and $\delta\eta_{zx}<\omega_a$, corresponds to the SP. Region~III, characterized by $\delta\eta_{zy}\geq\omega_a$, corresponds to the deformed phase, where the stationary point configuration is further rearranged.}
\end{figure}

As shown in Fig.~\ref{fig.2}, the fixed point configurations divide the parameter space into three regions associated with distinct phases. Region~I corresponds to the normal phase, where the polar fixed points in Eq.~(\ref{Point1}) consist of a north-pole local maximum and a south-pole global minimum. Region~II, defined by $f_{zx}\geq1$ and $\delta\eta_{zx}<\omega_a$, corresponds to the conventional superradiant phase: the south-pole minimum becomes a saddle point, while the two degenerate fixed points in Eq.~(\ref{Point2}) emerge as global minima. Region~III, characterized by $\delta \eta_{zy} \geq \omega_a$, is identified as a deformed phase, where the south-pole saddle turns into a local maximum and the two additional degenerate saddle points in Eq.~(\ref{Point3}) appear. The rearrangement of semiclassical stationary points establishes the ground-state phase diagram of the CHS model. The transition from a single global minimum to two degenerate minima signals the emergence of an intermediate parameter regime distinct from both the normal and the superradiant phases. For $\gamma=0$, this regime corresponds to an $XY$-type phase characterized by quasi-long-range order~\cite{10.1103/z8gv7yyk,PhysRevA.104.013303}. When $\gamma\neq0$, the continuous $U(1)$ symmetry is explicitly broken, and the intermediate regime evolves into a deformed phase with broken symmetry.
\begin{figure*}[htbp]
\centering
\includegraphics[width=0.85\textwidth]{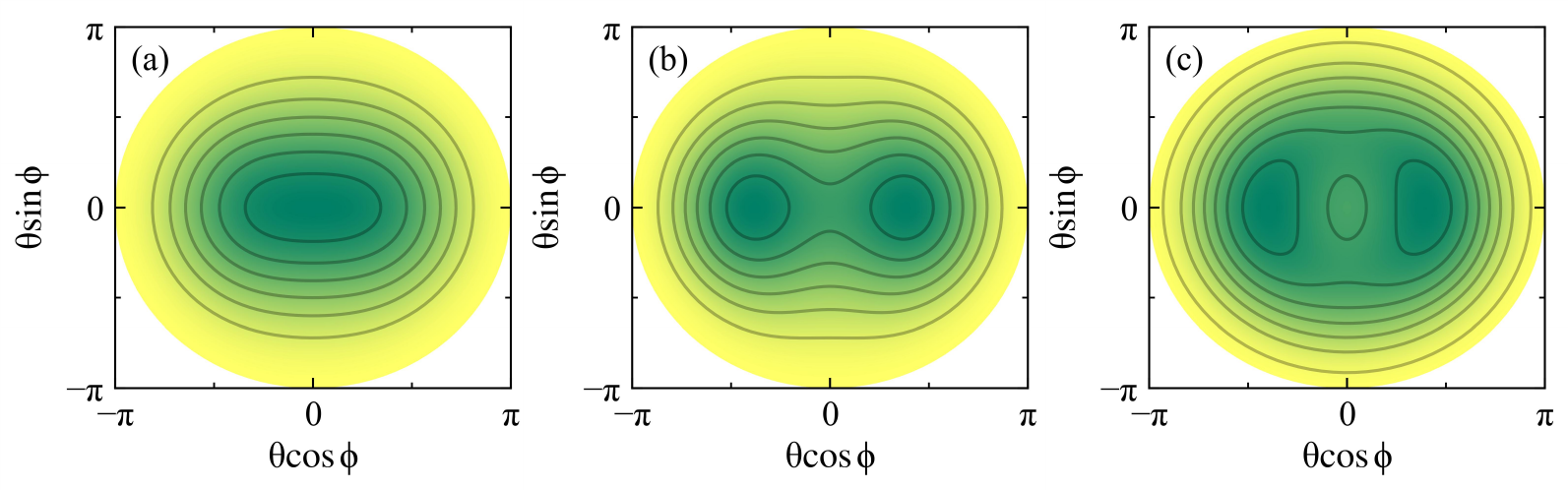}
\caption{\label{fig.3}Contour plots of the classical energy surface under resonance conditions \(\omega_a = \omega_c = 1\), represented in terms of the angular variables of the pseudospin \(j\). The azimuthal angle is denoted by \(\phi\), and the polar angle \(\theta\) is measured from the south pole, satisfying \( j_z = -j \cos\theta \).
(a) region I: \( g_1 = 0.1 \), \( g_2 = 0.5 \), \( \gamma = -0.2 \), \( \Delta = 1 \);
(b) region II: \( f_{zx} \geq 1 \), \( \delta \eta_{zx} < \omega_a \), with parameters \( g_1 = 0.8 \), \( g_2 = 0.5 \), \( \gamma = -0.2 \), \( \Delta = 1 \);
(c) region III: \( \delta \eta_{zy} \geq \omega_a \), with parameters \( g_1 = 0.5 \), \( g_2 = 0.5 \), \( \gamma = -0.2 \), \( \Delta = 3 \).}
\end{figure*}


This richer semiclassical structure motivates an analysis of the associated classical energy landscape. The minimum energy in each region is obtained by evaluating the classical Hamiltonian at the corresponding stable fixed points:
\begin{widetext}
\begin{equation}\label{E}
\mathcal{E}_{\min}
\equiv
\frac{E_{\min}}{\omega_a j}
=
\begin{cases}
-1+g_2\Delta,
&
\text{otherwise},
\quad \text{(I)}, \\[6pt]
-\dfrac{f_{zx}+f_{zx}^{-1}}{2}+g_2\Delta,
&
f_{zx}\geq1~\text{and}~\delta\eta_{zx}<\omega_a,
\quad \text{(II)}, \\[8pt]
-\dfrac{f_{zy}+f_{zy}^{-1}}{2}+g_2\Delta,
&
\delta \eta_{zy} \geq \omega_a,
\quad \text{(III)}.
\end{cases}
\end{equation}
\end{widetext}
Figure~\ref{fig.3} presents contour plots of the classical energy surfaces within the three regions defined by the fixed-point analysis. In region I, the system exhibits a local maximum with energy
\begin{equation}\label{E_max}
  \mathcal{E}_+ = 1 + g_2 \Delta,
\end{equation}
which lies above the global minimum \(\mathcal{E}_{x}\), 
Fig.~\ref{fig.3}(a). In region II, the system develops a saddle point with energy
\begin{equation}\label{E_saddle}
  \mathcal{E}_- = g_2 \Delta - 1,
\end{equation}
in addition to a local maximum at \(\mathcal{E}_+\). It can be readily verified that the energy levels satisfy
\(\mathcal{E}_{x} < \mathcal{E}_- < \mathcal{E}_+\) within this region [Fig.~\ref{fig.3}(b)]. In region III, a new saddle point appears with energy
\begin{equation}\label{E_s}
  \mathcal{E}_{y} \equiv - \frac{1}{2} (f_{zy} + f_{zy}^{-1}) + 2 g_2 \Delta,
\end{equation}
accompanied by two local maxima with energies \(\mathcal{E}_\pm\). The energy ordering in this region satisfies
\begin{equation}\label{E_order}
  \mathcal{E}_{x} < \mathcal{E}_{y} < \mathcal{E}_- < \mathcal{E}_+.
\end{equation}
The structure and ordering of these critical energy points have profound implications for the system's quantum behavior. In particular, the emergence of additional saddle points and degenerate minima alters the topology of the energy surface, which is directly reflected in the DoS. For instance, the presence of a saddle point typically leads to a nonanalyticity or divergence in the DoS, signaling the occurrence of an ESQPT.
The hierarchical structure $\mathcal{E}_{x}<\mathcal{E}_{y}<\mathcal{E}_{-}<\mathcal{E}_{+}$ in region III provides a distinctive hallmark of the deformed phase. Unlike the normal phase and the conventional superradiant phase, the deformed phase supports multiple characteristic critical energies originating from its richer stationary-point topology, thereby creating the most favorable regime for the emergence of multiple ESQPT signatures.
\section{EXCITED STATE QUANTUM PHASE TRANSITIONS} \label{section4}
ESQPTs are characterized by nonanalyticities in the DoS at critical energies, inducing singular features in the energy spectrum and anomalous nonequilibrium dynamics. Going beyond the conventional ground-state paradigm of QPTs, ESQPTs extend critical behavior into highly excited regimes. In particular, the deformed phase identified above provides a favorable setting for ESQPTs, since its richer stationary-point structure gives rise to multiple characteristic critical energies. In this section, we investigate the ESQPTs by combining spectral statistics with DoS analysis. 
\subsection{Energy spectrum}
\begin{figure*}[htbp]
\centering
\includegraphics[width=0.95\textwidth]{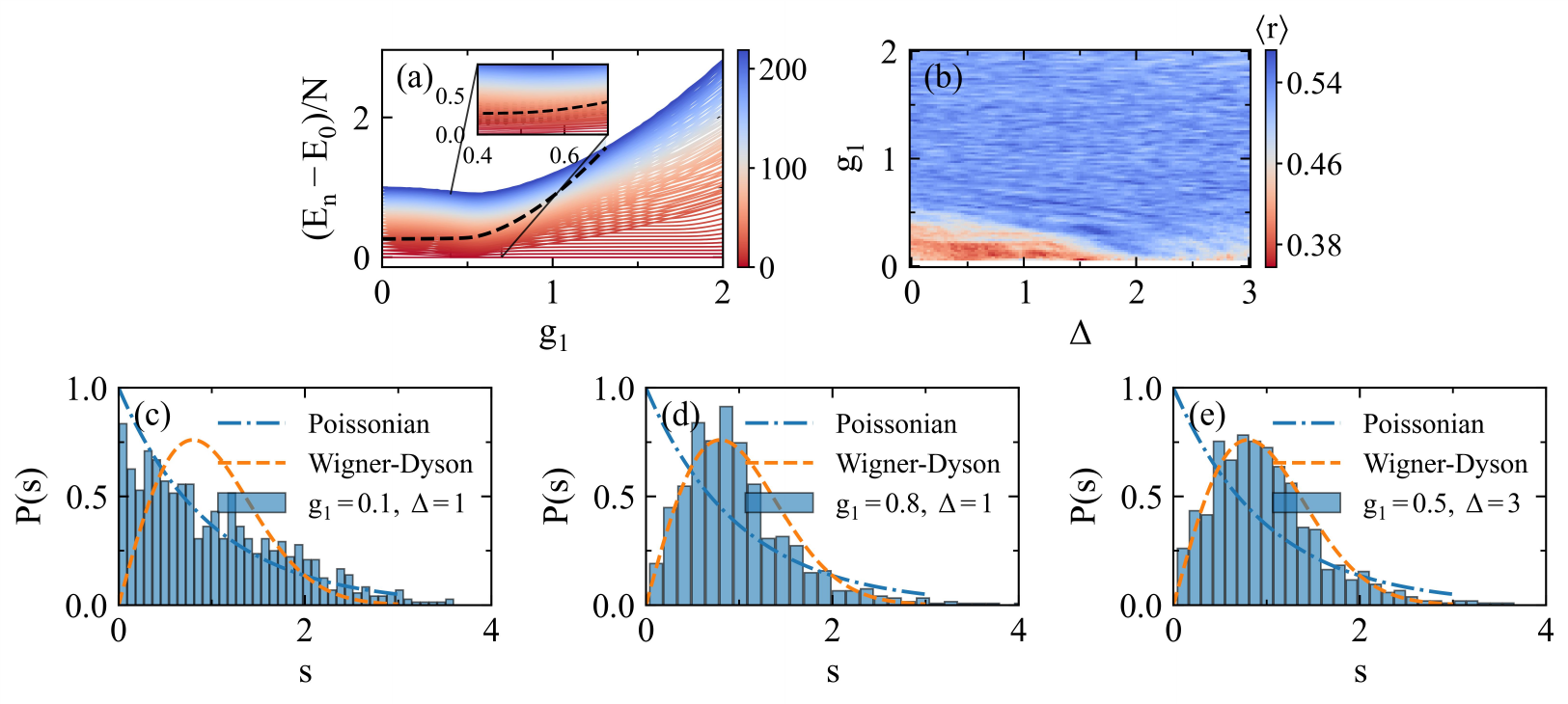}
\caption{\label{fig.4}Energy spectra and spectral statistics of the system.
(a) Excitation spectrum $(E_n - E_0)/N$ as a function of the spin--cavity coupling $g_1$ for $g_2=0.5$, $\gamma=-0.2$, and $\Delta=1$. The inset shows a magnified view of the critical energy region.
(b) Average adjacent-level spacing ratio $\langle r \rangle$ in the $(g_1,\Delta)$ parameter plane, computed from the lowest 1000 energy levels for $\omega_a=\omega_c=1$, $\gamma=-0.2$, and $N=20$.
The lower region is masked due to spurious large values induced by near-degenerate energy levels.
(c)--(e) Level-spacing distributions in different parameter regimes: (c) Poisson statistics in region~I; (d) Wigner--Dyson statistics in region~II; and (e) Wigner--Dyson statistics in region~III.
}
\end{figure*}
We first examine the excitation spectrum, which provides direct spectral signatures of the ESQPT. As displayed in Fig.~\ref{fig.4}(a), the spectrum undergoes a pronounced restructuring with increasing spin-cavity coupling $g_1$. For $g_1>g_{1c}=0.447$ (see also Eq. (\ref{g1c})), a finite-size precursor of the ESQPT becomes visible through the emergence of a distinct interface separating predominantly horizontal and tilted level contours. This qualitative rearrangement signals a change in the underlying classical energy landscape. The inset emphasizes the critical region, where the spectral reorganization near the semiclassical critical energy is resolved more clearly.

To characterize this restructuring quantitatively, we analyze the adjacent-level spacing ratio, which is widely used to diagnose regular and chaotic spectral behavior because it is largely insensitive to unfolding~\cite{PhysRevLett.110.084101,PhysRevLett.118.080601}.
For the ordered eigenenergies $\{E_n\}$, we define the level spacing as $s_n=E_{n+1}-E_n$ and the adjacent spacing ratio as
\begin{equation}
r_n=\frac{\min(s_{n-1},s_n)}{\max(s_{n-1},s_n)}.
\end{equation}
The averaged ratio $\langle r\rangle$, evaluated over the selected eigenlevels, provides a convenient indicator of the underlying spectral statistics.
Random-matrix theory predicts $\langle r\rangle\approx0.386$ for Poisson statistics and $\langle r\rangle_{\rm GOE}\approx0.5307$ for the Gaussian orthogonal ensemble (GOE)~\cite{PhysRevLett.110.084101}.
In the cavity--spin-chain model, $\langle r\rangle$ remains close to the Poisson value for $g_1<g_{1c}$, indicating predominantly regular or nonergodic spectral behavior.
By contrast, for $g_1>g_{1c}$, $\langle r\rangle$ approaches the GOE benchmark over broad parameter regions, signaling the onset of chaotic and ergodic behavior.

This tendency is fully consistent with the level-spacing distributions shown in Figs.~\ref{fig.4}(c)--\ref{fig.4}(e).
After unfolding the spectrum to unit mean level density, nonergodic regimes follow the Poisson form $P(s)=e^{-s}$, whereas ergodic regimes exhibit the GOE Wigner--Dyson distribution, $P(s)=\frac{\pi}{2}s\exp[-(\pi/4)s^2]$, in agreement with the Berry--Tabor and Bohigas--Giannoni--Schmit conjectures, respectively~\cite{berry1977level,PhysRevLett.52.1,gubin2012quantum,PhysRevLett.118.080601}.
Specifically, for $\omega_a=\omega_c=1$ and $j=10$ ($N=20$), the system displays nonergodic statistics at $(g_1,\Delta)=(0.1,1)$~\cite{PhysRevE.50.888}$,$ whereas ergodic statistics emerge at $(0.8,1)$ and remain robust at $(0.5,3)$.
We note that in some parameter patches the numerically obtained $\langle r\rangle$ may slightly exceed $\langle r\rangle_{\rm GOE}$.
Such an overshoot is a common finite-size effect and is typically caused by nonuniversal structures, such as near-degenerate levels, limited statistics within a restricted energy window, or residual symmetry effects, rather than by a genuine violation of GOE universality.
Therefore, the crossover from Poisson-like to Wigner--Dyson statistics should be assessed jointly from $\langle r\rangle$ and the full distribution $P(s)$.

The global spectral structure and its singular features not only distinguish different quantum phases, but also leave characteristic fingerprints in the DoS.
To further characterize these spectral features and identify signatures of criticality, we now turn to the DoS.
\subsection{Density of states}
The critical energy can be determined by analytically evaluating the DoS of the classical model using Weyl's law~\cite{gutzwiller2013chaos}:
\begin{equation}\label{V_DoS}
\nu(E) = \frac{1}{(2\pi)^2} \int dq \, dp \, d\phi \, dj_z \, \delta\left(E - H_{\text{cl}}(q,~p,~\phi,~j_z)\right),
\end{equation}
which gives the volume of the accessible phase space at a given energy \(E\).

The integration over the bosonic canonical variables \((q,~p)\) yields a constant factor of \(2\pi / \omega_c\). The integration over the pseudospin sector is constrained by the inequality:
\begin{equation}
\begin{split}
\cos^2 \phi \geq \left( \frac{2}{1 - j_z^2} ( \frac{\eta_z}{2\omega_a} j_z^2 + j_z - \mathcal{E}) + \frac{\eta_y}{\omega_a} \right)
( \frac{g_1^2}{\gamma_c^2} - \frac{\eta_x}{\omega_a} + \frac{\eta_y}{\omega_a}),
\end{split}
\end{equation}
where \( \gamma_c = \sqrt{\omega_a \omega_c} / 2 \), and \(\phi\) satisfies the following expression:
\begin{equation}
\smash[b]{%
\begin{aligned}
\phi(j_z,\mathcal{E}) = \arccos \Bigg\{ &
\left[ \frac{2}{1 - j_z^2} \left( \frac{\eta_z}{2\omega_a} j_z^2 + j_z - \mathcal{E} \right) + \frac{\eta_y}{\omega_a} \right]^{1/2} \\
& \hspace{0.1cm} \times \left[ \frac{g_1^2}{\gamma_c^2} - \left( \frac{\eta_x}{\omega_a} - \frac{\eta_y}{\omega_a} \right) \right]^{-1/2}
\Bigg\}.
\end{aligned}
}
\end{equation}
\vskip 1.9em
\noindent
The allowed range of \(j_z\) is bounded by the conditions \( j_z^{(-)} < j_z^{(+)} \) and \( j_z^{(1)} < j_z^{(2)} \), where
\begin{equation}
\begin{aligned}
j_{z}^{(\pm)}(\varepsilon) &= -\frac{1}{f_{zx}} \left[ 1 \mp \sqrt{2 f_{zx} (\varepsilon - \varepsilon_{sx})} \right], \\
j_{z}^{(1,2)}(\varepsilon) &= -\frac{1}{f_{zy}} \left[ 1 \mp \sqrt{2 f_{zy} (\varepsilon - \varepsilon_{sy})} \right].
\end{aligned}
\end{equation}
\begin{figure*}[htbp]
\centering
\includegraphics[width=0.95\textwidth]{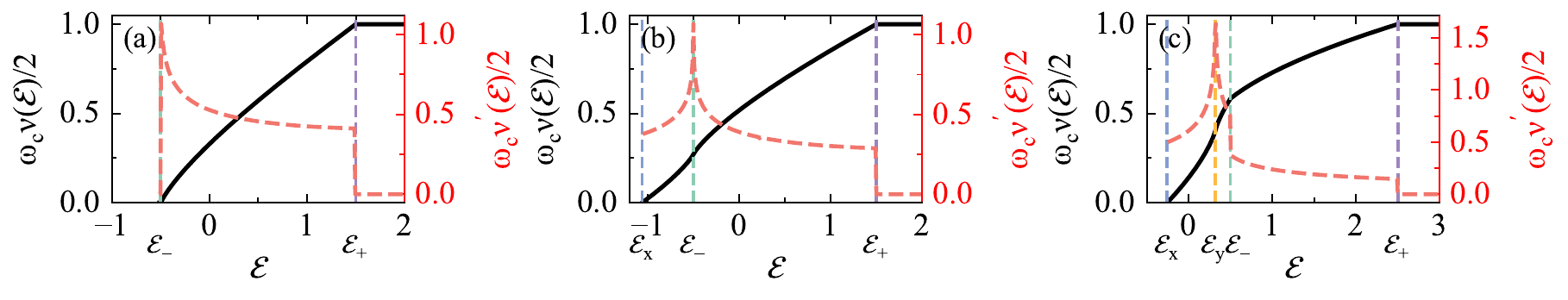}
\caption{\label{fig.5}The semiclassical DoS, \( \omega_c \nu(\mathcal{E}) / 2 \) (black dashed line), and its first derivative (red solid line) as functions of the scaled energy \( \mathcal{E} \equiv E / (j \omega_a) \), with parameters identical to those of Fig.~\ref{fig.3}.}
\end{figure*}

In region I, the semiclassical DoS \(\omega_c \nu(\mathcal{E}) / 2\) splits into two subregions:
\begin{widetext}
\begin{equation}
\frac{\omega_c}{2} \nu(\mathcal{E}) =
\begin{cases}
\frac{1}{\pi} \int_{j_{z}^{(1)}}^{j_{z}^{(+)}} \phi(j_z, \mathcal{E}) \, dj_z + \frac{1}{2} \left( j_{z}^{(1)} + 1 \right), & \mathcal{E} \in [\mathcal{E}_{-}, \mathcal{E}_{+}], \, \gamma \in [0, \infty), \\
1, & \mathcal{E} > \mathcal{E}_{+}, \, \gamma \in [0, \infty).
\end{cases}
\end{equation}
\end{widetext}
Its first derivative exhibits a jump-type discontinuity at the critical energy \(\mathcal{E}_{+}\), as shown in Fig.~\ref{fig.5}~(a), signaling a semiclassical signature of an ESQPT.

In region II, the DoS consists of three distinct subregions:
\begin{widetext}
\begin{equation}
\frac{\omega_c}{2} \nu(\mathcal{E}) =
\begin{cases}
\frac{1}{\pi} \int_{j_{z}^{(-)}}^{j_{z}^{(+)}} \phi(j_z,~\mathcal{E}) \, dj_z,~& \mathcal{E} \in [\mathcal{E}_{x},~\mathcal{E}_{-}],~\,~ g_1 \in [g_1^c,~\infty),~\\
\frac{1}{\pi} \int_{j_{z}^{(1)}}^{j_{z}^{(+)}} \phi(j_z,~\mathcal{E}) \, dj_z + \frac{1}{2} \left( j_{z}^{(1)} + 1 \right),~& \mathcal{E} \in (\mathcal{E}_{-},~\mathcal{E}_{+}],~\,~\gamma \in [0,~\infty), \\
1, & \mathcal{E} > \mathcal{E}_{+},~\,~\gamma \in [0,~\infty).
\end{cases}
\end{equation}
\end{widetext}
At \(\mathcal{E}_{-}\), the first derivative of the DoS shows a logarithmic discontinuity, while a jump discontinuity persists at \(\mathcal{E}_{+}\), as illustrated in Fig.~\ref{fig.5}(b). This behavior, characteristic of the Dicke model, indicates the presence of two qualitatively different ESQPTs at energies \(\mathcal{E}_{\pm}\).

In region III, for \(\delta \eta_{zy} \geq \omega_a\), the system exhibits distinct behavior compared to the standard Dicke model, resulting in four DoS subregions:
\begin{widetext}
\begin{equation}
\frac{\omega_c}{2} \nu(\mathcal{E}) =
\begin{cases}
\frac{1}{\pi} \int_{j_{z}^{(-)}}^{j_{z}^{(+)}} \phi(j_z,~\mathcal{E}) \, dj_z,~& \mathcal{E} \in [\mathcal{E}_{x},~\mathcal{E}_{y}],~\,~ g_1 \in [g_1^c,~\infty), \\
\frac{1}{\pi} \left[ \int_{j_{z}^{(-)}}^{j_{z}^{(1)}} \phi(j_z, \mathcal{E}) \, dj_z + \int_{j_{z}^{(2)}}^{j_{z}^{(+)}} \phi(j_z,~ \mathcal{E}) \, dj_z \right] + \frac{1}{2} \left( j_{z}^{(2)} - j_{z}^{(1)} \right), & \mathcal{E} \in (\mathcal{E}_{y},~\mathcal{E}_{-}],~ g_1 \in [g_1^c,~\infty),~\,\\
&\quad \Delta \eta_{zy} \geq \omega_a,~\\
\frac{1}{\pi} \int_{j_{z}^{(1)}}^{j_{z}^{(+)}} \phi(j_z,~\mathcal{E}) \, dj_z + \frac{1}{2} \left( j_{z}^{(1)} + 1 \right),~& \mathcal{E} \in (\mathcal{E}_{-},~\mathcal{E}_{+}], \\
1, & \mathcal{E} > \mathcal{E}_{+}.
\end{cases}
\end{equation}
\end{widetext}
The logarithmic discontinuity in the DoS derivative is now relocated to the critical energy \(\mathcal{E}_{y}\), which corresponds to a newly emerged saddle point on the energy surface. In addition, a new jump discontinuity appears at \(\mathcal{E}_{-}\), while the existing discontinuity at \(\mathcal{E}_{+}\) remains. These features suggest the presence of three distinct types of ESQPTs in region III, as shown in Fig.~\ref{fig.5}(c).

Peres lattices---scatter plots of an observable's expectation value versus energy---provide a compact, qualitative probe of spectral organization~\cite{Peres1984}. In integrable regimes, the points assemble into regular grids (consistent with Einstein, Brillouin, and Keller quantization), whereas nonintegrable perturbations progressively deteriorate the lattice into diffuse patches as chaos develops~\cite{PhysRevA.89.032102}. The construction is independent of the choice of Peres operator (any Hermitian observable may be used) and, by resolving structure at fixed energy, complements bulk indicators such as level-spacing statistics.

\begin{figure*}[htbp]
\centering
\includegraphics[width=0.95\textwidth]{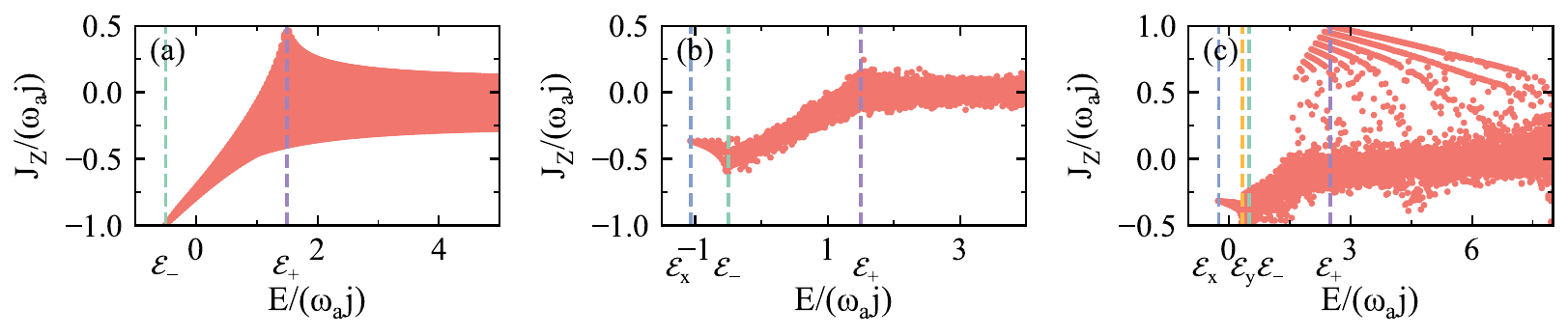}
\caption{\label{fig.6}Peres lattices for the CHS model: the scaled spin projection \(\langle J_z\rangle/(\omega_a j)\) versus the scaled energy \(\mathcal{E}\equiv E/(\omega_a j)\). Panels (a)--(c) correspond to regions I--III of Fig.~\ref{fig.3}, using the same parameters. Vertical dashed lines indicate the semiclassical energies \(\mathcal{E}_{-}\), \(\mathcal{E}_{+}\), \(\mathcal{E}_{x}\) and \(\mathcal{E}_{y}\) from the fixed-point analysis.}
\end{figure*}
For systems exhibiting ESQPTs, Peres lattices visualize critical energies via abrupt reorganizations of the point cloud and the emergence of boundary-like ridge structures~\cite{PhysRevA.89.032101,PhysRevA.89.032102}. In the CHS model, lattices of \(\langle J_z\rangle/(\omega_a j)\) versus the scaled energy \(\mathcal{E}\) (see Fig.~\ref{fig.6}) display precursors of a static ESQPT near \(\mathcal{E}_+\), evident in panels (a)--(c), and specifically in regions~II and~III features of a dynamic ESQPT near \(\mathcal{E}_-\) [Fig.~\ref{fig.6}(b,c)]. Notably, in region~III a partial revival of order appears at large spin projections near \(\mathcal{E}_+\) [Fig.~\ref{fig.6}(c)], echoing observations in extended Dicke models where strong nonlinear couplings restore regular islands around ESQPT energies~\cite{PhysRevA.98.043805}. Used alongside density-of-states singularities and level statistics, these lattices offer a sensitive, low-overhead diagnostic of excited-state criticality and its interplay with the integrable--chaotic crossover in hybrid light--matter systems~\cite{PhysRevA.89.032102,PhysRevA.98.043805}.
\section{ERGODIC-NONERGODIC TRANSITION} \label{section5}
Ergodicity breaking in quantum many-body systems has emerged as a central theme in understanding nonthermal behavior beyond conventional statistical mechanics. In particular, the transition between ergodic and nonergodic extended phases offers crucial insight into the interplay between spectral properties and wavefunction structure. In this section, we examine the ENET in the CHS model by analyzing both static eigenstate features and post-quench dynamics. These results reveal characteristic signatures of ergodicity breaking and a tight connection to excited-state quantum criticality, as evidenced by the behavior of participation ratios and multifractal dimensions.
\begin{figure*}[htpb]
\centering
\includegraphics[width=0.95\textwidth]{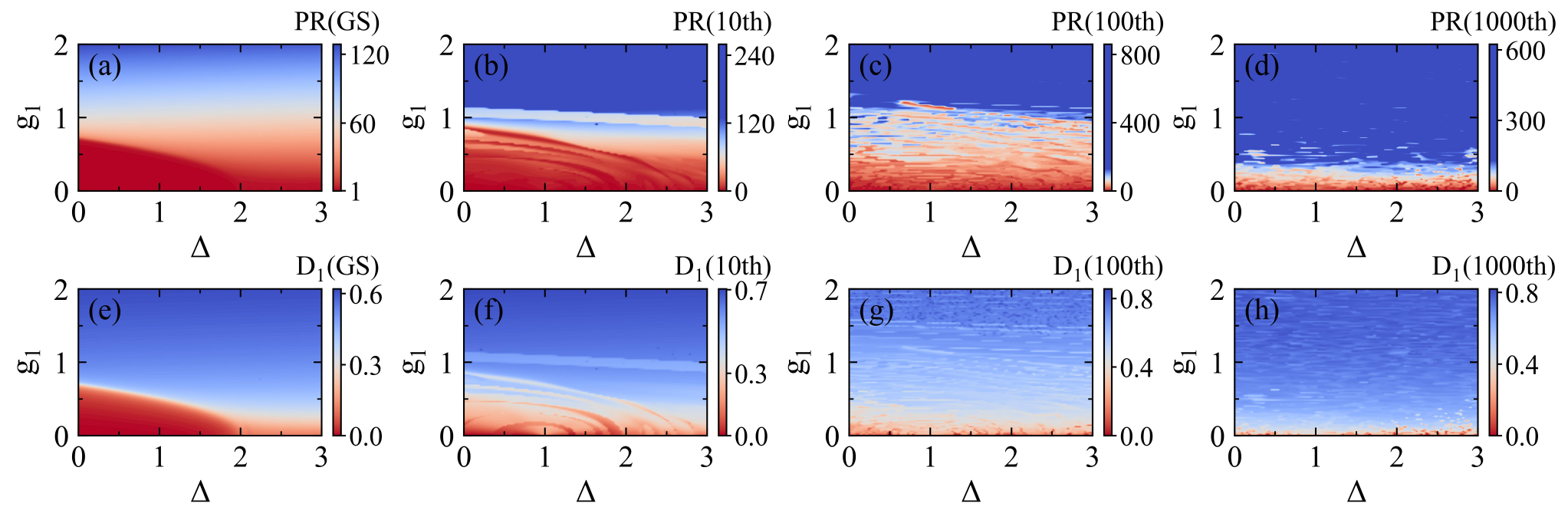}
\caption{\label{fig.7}Phase diagrams of the CHS model showing the participation ratio and multifractal dimension \( D_1 \) as functions of the coupling parameters \( g_1 \) and \( \Delta \), evaluated for representative eigenstates.
(a)--(d) PR maps corresponding to the ground state, the 10th (low-lying), 100th (intermediate), and 1000th (highly excited) states, respectively.
(e) The ground-state result of \( D_1 \), indicating a transition from a localized regime (NP) to a multifractal regime (SP).
(f),~(g) \( D_1 \) distributions for the 10th and 100th excited states.
(h) A transition in the 1000th excited state from a non-ergodic extended phase (multifractal) to an ergodic delocalized phase.
All results are obtained with fixed parameters \( \omega_a = \omega_c = 1, \gamma = -0.2, N = 20 \), and \( n_{\max} = 200 \).}
\end{figure*}
\subsection{Static indicators}
Different eigenstates play distinct roles in various types of quantum phase transitions. The ground state primarily reflects the transition from the normal phase to the superradiant phase, or equivalently, from a localized to a multifractal phase. In contrast, intermediate excited states are often associated with transitions between non-ergodic and ergodic behavior. In this section, we investigate the phase diagram of the CHS model in the \((g_1,~\Delta)\) parameter plane, focusing on the behavior of different eigenstates. To this end, we employ the participation ratio (PR) and multifractal dimensions to characterize the localization properties of the eigenstates. Particular attention is paid to the emergence of multifractal features in the excited-state spectrum.
\begin{figure*}[t]
\centering
\includegraphics[width=0.95\textwidth]{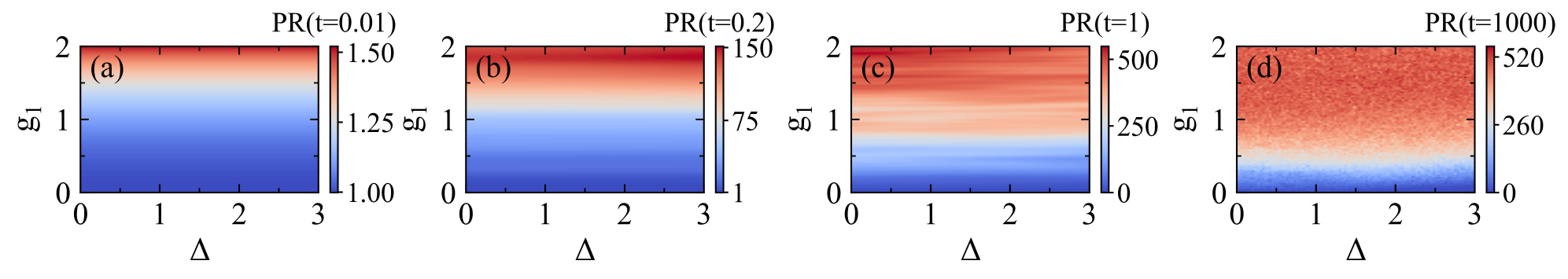}
\caption{\label{fig.8}PR following a quantum quench in the cavity-coupled Heisenberg spin-1/2 chain, shown at different evolution times \( t \). The system parameters are \( \omega_a = \omega_c = 1, \gamma = -0.2, g_2 = 0.5\), spin number \( N = 20 \), and bosonic truncation dimension \( n_{\max} = 100 \). The panels correspond to: (a)~\( t = 0.01 \), (b)~\( t = 0.2 \), (c)~\( t = 1 \), and (d)~\( t = 1000 \). The initial state is chosen as an intermediate excited eigenstate of the decoupled Hamiltonian \( H_0 \). Time is expressed in units of \( \omega_a^{-1} \).}
\end{figure*}
The PR provides a quantitative measure of the degree of localization or delocalization of a quantum state. Beyond its direct definition, its scaling with the Hilbert-space dimension $N_D$ offers further insight into the structure of quantum eigenstates. For a finite system, we use the estimator of the generalized fractal dimension~\cite{PhysRevLett.123.180601,PhysRevLett.122.106603,PhysRevB.109.235432}
\begin{align}\label{Dq}
D_q=\frac{1}{1-q}\frac{\ln\!\left(\sum_{j=1}^{N_D}|\psi_j|^{2q}\right)}{\ln N_D},
\end{align}
where $|\psi\rangle$ is an eigenstate of the Hamiltonian. The quantity
\(S_q=\frac{1}{1-q}\ln\!\left(\sum_{j=1}^{N_D}|\psi_j|^{2q}\right)\) is the $q$-dependent participation entropy. In the Shannon limit $q\to1$, one obtains \(S_1=-\sum_{j=1}^{N_D}|\psi_j|^2\ln |\psi_j|^2\), while for $q=2$ it is directly related to the conventional participation ratio through
\(S_2=\ln(\mathrm{PR})\). For a fully delocalized state, $S_q\approx \ln N_D$, so that $D_q\approx 1$. By contrast, for a fully localized state, $S_q$ remains finite and $D_q\approx 0$. Between these two limits, an intermediate regime may arise in which the wave function is extended but nonergodic in the chosen basis. In that case, $S_q$ scales as $D_q\ln N_D$ with $0<D_q<1$, indicating multifractal behavior.

For $N=20$ and $n_{\max}=200$, the phase maps of the PR and the information (Shannon) dimension $D_1$ reveal how localization and ergodicity evolve across the CHS parameter space [Fig.~\ref{fig.7}]. For the ground state, a normal-to-superradiant transition appears along the semiclassical boundary $f_{zx}=1$: in the normal phase the PR is small and $D_1\simeq0$, whereas in the superradiant phase the PR increases and $0<D_1<1$, indicating the emergence of multifractal eigenstates. With increasing excitation energy, the dominant critical behavior evolves from the normal--superradiant transition to a crossover between nonergodic and ergodic extended states. For the intermediate excited state [panel~(c)], the PR remains low in the nonergodic region but increases markedly in the ergodic region. The corresponding $D_1$ map [panel~(g)] shows a consistent evolution: red regions with $0<D_1<1$ correspond to multifractal, nonergodic extended states, whereas blue regions with $D_1\approx0.9$ correspond to strongly extended, nearly ergodic states. For the highly excited state [panels~(d) and (h)], the same qualitative trend persists, although both the PR and the corresponding $D_1$ in the ergodic region are moderately reduced compared with those of the intermediate excited state. Taken together, the PR and $D_1$ distinguish the ground-state normal-to-superradiant transition, accompanied by the evolution from localized to multifractal eigenstates, from the crossover between nonergodic and ergodic extended states at higher excitation energies.
\subsection{Quench dynamics across the ESQPT}
\begin{figure*}[htpb]
\centering
\includegraphics[width=0.9\textwidth]{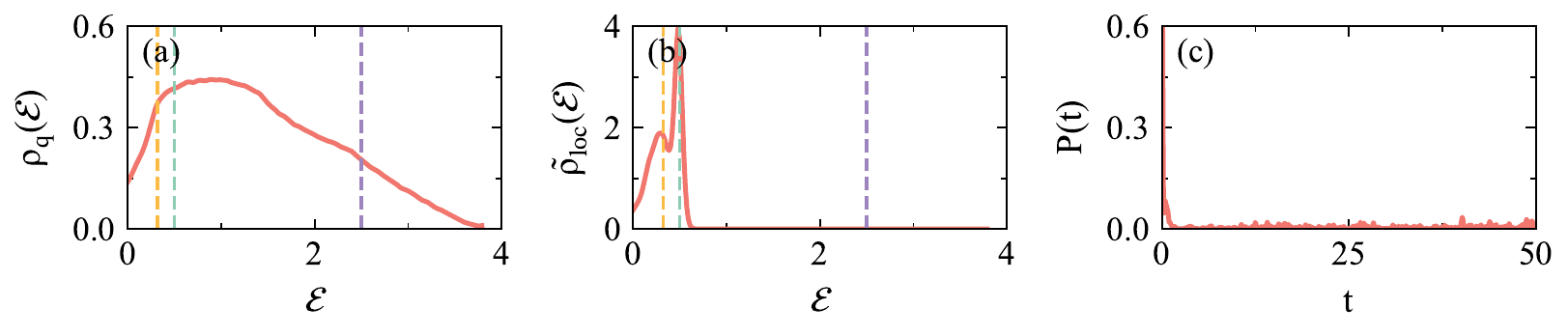}
\caption{\label{fig.9}
Signatures of the ESQPT in quench dynamics from Region I to Region III.
(a) Smoothed density of states $\rho(\mathcal{E})$ of the post-quench Hamiltonian.
(b) Smoothed local density of states $\tilde{\rho}_{\rm loc}(\mathcal{E})$ of the initial state projected onto the eigenstates of the post-quench Hamiltonian.
(c) Survival probability $P(t)$ after the quench.
}
\end{figure*}
To investigate the quench dynamics of the closed CHS model, we prepare the system in an eigenstate of the noninteracting pre-quench Hamiltonian $H_0 = \omega_c a^\dagger a + \omega_aJ_z$. At $t=0$, the Hamiltonian is suddenly quenched to $H = H_0 + H_1$, where $H_1 = \frac{\omega_a g_2}{N} \Big[ J_+ J_- + J_- J_+ + \gamma (J_+^2 + J_-^2) + 2\Delta J_z^2 - \frac{N}{2}(2 + \Delta)\Big] + \frac{g_1}{\sqrt{N}} (J_+ + J_-) (a^\dagger + a)$, thereby simultaneously switching on both the spin--spin interaction and the spin--cavity coupling. The post-quench state evolves unitarily according to \(|\psi(t)\rangle = e^{-iHt}|\psi_{\rm in}\rangle=\sum_{\alpha}C_{\alpha}(t)|\alpha\rangle,\)
where $|\alpha\rangle=|n,j,m\rangle$ denotes the computational basis and $C_{\alpha}(t)$ is the corresponding expansion coefficient. To characterize the spreading of the wave function over the Hilbert space, we calculate the time-dependent participation ratio, \(\mathrm{PR}(t)=\left[\sum_{\alpha}|C_{\alpha}(t)|^4\right]^{-1},\) at representative times $t=0.01$, $0.2$, $1$, and $1000$. Figure~\ref{fig.8} displays the corresponding PR in the $(g_1,\Delta)$ parameter plane. At early times [$t=0.01$, Fig.~\ref{fig.8}(a)], the PR remains low throughout the phase diagram, indicating that the dynamics is still dominated by the initial state. By $t=0.2$ [Fig.~\ref{fig.8}(b)], the PR begins to increase over part of the parameter space, signaling the onset of wave-function delocalization. At $t=1$ [Fig.~\ref{fig.8}(c)], the PR reaches large values over extended parameter regions, indicating substantial coherent spreading and nearly ergodic behavior. At long times, the contrast between different dynamical regimes becomes more pronounced: in nonergodic regions, the dynamics remains confined to a finite effective subspace and the PR stays low, whereas in ergodic regions, the PR remains high, reflecting sustained delocalization. These results provide evidence for an ergodic-nonergodic transition in the CHS model.

While the PR captures the global spreading of the wave function in the computational basis, it does not directly reveal how the spectral properties of the post-quench Hamiltonian control the subsequent dynamics. To gain this complementary perspective, we expand the initial state in the eigenbasis of the quenched Hamiltonian,
\begin{equation}\label{psi}
|\psi_{\rm in}\rangle=\sum_n c_n |E_n\rangle ,
\end{equation}
where $|E_n\rangle$ are the eigenstates of $H$. The coefficients $|c_n|^2$ determine the spectral weights of the local density of states (local DoS),
\(\rho_{\rm loc}(E)=\sum_n |c_n|^2 \delta(E-E_n)\), which encodes how the initial state is distributed over the post-quench spectrum~\cite{PhysRevA.94.012113}. To elucidate the dynamical role of the ESQPT, we consider quenches from Region~I to Region~III, where the semiclassical DoS exhibits a critical energy associated with a qualitative change in the underlying phase-space structure. The initial state is chosen such that its mean post-quench energy lies close to this critical point, allowing the dynamics to probe the spectral sector governed by the ESQPT. Figure~\ref{fig.9} displays the smoothed density of states of the post-quench Hamiltonian, the smoothed local DoS of the selected initial state, and the corresponding survival probability,
\begin{equation}\label{Pt}
P(t)=|\langle \psi_{\rm in}|e^{-iHt}|\psi_{\rm in}\rangle|^2 .
\end{equation}
The local DoS is found to concentrate near the ESQPT critical energy identified in Fig.~\ref{fig.5}, reflecting the ESQPT-induced restructuring of the eigenstates. This concentration indicates that the quench dynamics is dominated by the corresponding spectral sector, leading to relaxation behavior that deviates markedly from simple exponential decay~\cite{PhysRevA.83.033802}. These results show that the ESQPT leaves clear dynamical imprints on quench processes through its reorganization of the spectral structure in energy space. Together with the participation-ratio analysis, the local-DoS approach establishes a direct connection between critical spectral properties and nonequilibrium quantum dynamics in the CHS model.
\section{CONCLUSIONS} \label{section6}
We have investigated QPTs, ESQPTs, and the ENET in the CHS model with spin-spin interactions through complementary semiclassical, spectral, and dynamical analyses. In the ground-state regime, beyond the NP and SP, we have found an intermediate deformed phase. These three phases are separated by the auxiliary function \(f_{z\alpha}=1\): in the NP, one has \(|\langle J_{z}\rangle|=1\), whereas in both the SP and the deformed phase \(|\langle J_{z}\rangle|\in(0,1)\). We have also observed the scaled photon number satisfies \(\langle \hat{a}^\dagger \hat{a}\rangle/j \simeq 0\) in the NP, while becoming finite in the SP and deformed phases, indicating the onset of macroscopic bosonic excitations in hybrid light-matter systems. In the excited-state regime, we have identified ESQPTs as a steplike enhancement of the DoS near a critical energy, and have found that level-statistics diagnostics the adjacent-gap ratio and level-spacing distribution---switch between Poisson and Wigner-Dyson behavior with variations in the light-matter coupling and spin-spin interaction strength, thereby marking the transition from nonergodic to ergodic regimes. Our semiclassical analysis has further predicted a DoS discontinuity at the critical energy, which acts as a definitive precursor of ESQPTs and establishes a direct semiclassical quantum correspondence for excited-state criticality. Via the PR and an energy-resolved spectral characterization of quantum-quench dynamics, we have explicitly demonstrated that ESQPT induced spectral restructuring governs the system's nonequilibrium time evolution and leaves clear dynamical fingerprints in the local DoS and survival probability, thereby establishing a direct link between excited-state criticality and nonequilibrium dynamics. Our results demonstrate that spin-spin interactions and light-matter coupling jointly govern the critical properties of such quantum many-body systems across both ground and excited states. We have provided a unified physical picture connecting ground-state phase structure, excited-state spectral characteristics, and nonequilibrium dynamics. These findings resolve key ambiguities in phase classification of the spin-included CHS model, reveal the microscopic mechanism through which ESQPTs control nonequilibrium behavior, and offer a consistent theoretical framework for studying critical phenomena in light--matter interacting many-body systems.
\section*{Acknowledgments}
The work is supported by the National Natural Science Foundation of China (Grant No. 12475026) and the Natural Science Foundation of Gansu Province (No. 25JRRA799).

\bibliography{reference}
\end{document}